\documentclass[cits]{PoS}
\usepackage{bm}
\newcommand{\be}{\begin{equation}}
\newcommand{\ee}{\end{equation}}
\newcommand{\bea}{\begin{eqnarray}}
\newcommand{\eea}{\end{eqnarray}}
\newcommand{\Slash}[1]{{\ooalign{\hfil/\hfil\crcr$#1$}}}

\title{OPE for B-meson distribution amplitude and dimension-5 HQET operators}

\ShortTitle{OPE for B-meson distribution amplitude and dimension-5 HQET operators}

\author{Hiroyuki Kawamura%
\thanks{Supported in part by the UK Science \& 
Technology Facilities Council under grant number PP/E007414/1.}
\\
 Department of Mathematical Sciences, University of Liverpool, Liverpool, L69 3BX, United Kingdom
\\
        E-mail: \email{Hiroyuki.Kawamura@liverpool.ac.uk}}

\author{\speaker{Kazuhiro Tanaka}%
\thanks{Supported by the Grant-in-Aid 
for Scientific Research No.~B-19340063.}
\\
Department of Physics, Juntendo University,
    Inba, Chiba 270-1695, Japan
\\
        E-mail: \email{tanakak@sakura.juntendo.ac.jp}}

\abstract{
The $B$-meson light-cone distribution amplitude (LCDA) is
defined as the matrix element 
of a quark-antiquark bilocal light-cone operator in the heavy-quark effective theory (HQET)
and is a building block
of QCD factorization formula for exclusive $B$-meson decays.
When the corresponding bilocal HQET operator 
has a light-like distance $t$ between the quark and antiquark fields, 
the scale $\sim 1/t$ separates the UV and IR regions,
which induce the cusp singularity in 
radiative corrections and the mixing of multiparticle states 
in nonperturbative corrections, respectively.
We treat
the bilocal HQET operator 
based on the operator product expansion (OPE), 
disentangling the singularities from the IR and UV regions systematically.
The matching at the next-to-leading order $\alpha_s$ is performed 
in the $\overline{\rm MS}$ scheme with a complete set of local operators of dimension $d \le 5$, 
through a manifestly gauge-invariant calculation 
organizing all contributions in the coordinate space.
The result exhibits
the Wilson coefficients
with Sudakov-type double logarithms
and 
the higher-dimensional operators with additional 
gluons.
This OPE yields
the $B$-meson LCDA for $t$ less than $\sim 1$ GeV$^{-1}$, in
terms of $\bar{\Lambda}= m_B - m_b$
and the two 
additional HQET parameters as matrix elements of dimension-5 operators.
The impact of these novel HQET parameters on the integral relevant to exclusive 
$B$ decays, $\lambda_B$, is also discussed.
}

\FullConference{International Workshop on Effective Field Theories: from the pion to the upsilon \\
		February 2-6 2009\\
		Valencia, Spain}

\begin{document}

For the exclusive $B$-meson decays, such as $B \to \pi \pi$, 
$\rho \gamma, \ldots$, systematic methods have been developed
using QCD factorization based on the heavy-quark 
limit~\cite{Beneke:2000ry,Bauer:2001cu,Bell08}.
In the corresponding factorization formula of the decay amplitude,
essential roles are played by
the light-cone 
distribution amplitudes (LCDAs) for the participating mesons, 
which include
nonperturbative long-distance contributions.
In particular, in addition to the LCDAs for the light mesons $\pi, \rho$, etc.,
produced in the final state,
the LCDA $\tilde{\phi}_+$ for the $B$ meson, 
defined
as the vacuum-to-meson matrix element~\cite{Grozin:1997pq},
\begin{equation}
\tilde{\phi}_+(t, \mu)
=
\frac{1}{iF(\mu)}
\langle 0|
\bar{q}(tn)
{\rm P}e^{ig\int_0^td\lambda n\cdot A(\lambda n)}
\Slash{n}
\gamma_5h_v(0) 
|\bar{B}(v)\rangle 
=\int d\omega e^{-i\omega t}
\phi_+(\omega, \mu)\ ,
\label{eq1}
\end{equation}
also participates in 
processes where large momentum is transferred to the soft 
spectator quark via hard gluon
exchange~\cite{Beneke:2000ry,Bauer:2001cu,Bell08}.
Here,
the bilocal operator 
is built of 
the $b$-quark and light-antiquark 
fields, $h_v(0)$ and $\bar{q}(tn)$, linked by the Wilson line
at a light-like 
separation 
$tn$, with $n_\mu$ as the light-like vector 
($n^2 =0$, $n\cdot v=1$), and $v_\mu$ representing the
4-velocity of the $B$ meson;
a difference between (\ref{eq1}) and the familiar pion-LCDA is that $h_v(0)$ 
is an effective field in the heavy-quark effective theory (HQET).
$\mu$ denotes
the scale where the 
operator is renormalized, and 
$F(\mu)$ is the decay constant in HQET, $F(\mu)=-i\langle 0|
\bar{q}
\Slash{n}
\gamma_5h_v 
|\bar{B}(v)\rangle$.
The RHS in (\ref{eq1}) defines 
the momentum representation, 
with $\omega v^+$ denoting the LC component of the momentum of the light antiquark.

The ``IR structure'' of 
(\ref{eq1}),
studied
using constraints from the equations of motion (EOM)
and heavy-quark symmetry~\cite{KKQT}, 
as well as the ``UV structure'',
calculated in the 1-loop renormalization
of 
the bilocal operator in 
(\ref{eq1})~\cite{Lange:2003ff},
is notoriously
peculiar compared with the pion LCDA.
For a full description of (\ref{eq1}) which would involve
a complicated mixture 
of the IR and UV structures,
we first calculate the radiative corrections, 
taking into account hard and soft/collinear loops.
The one-particle-irreducible 1-loop diagrams (1LDs) for the 2-point function
$\langle \bar{q}(tn)\Slash{n}\gamma_5h_v(0) \rangle$ 
of (\ref{eq1})
yield~\cite{KT09} 
($\langle \cdots \rangle \equiv \langle 0| \cdots|\bar{B}(v) \rangle$, 
the Wilson line 
is suppressed, 
and $C_F = (N_c^2-1)/(2N_c)$)
\begin{eqnarray}
&&
\!\!\!\!\!\!
{\rm 1LDs}=
\frac{\alpha_s C_F }{2\pi }
\int_0^1 {d\xi } \left[ \left\{  - \left( \frac{1}{2\varepsilon_{UV}^2} 
+ \frac{L}{\varepsilon_{UV}} + L^2+ \frac{5\pi^2 }{24} \right)
\delta (1 - \xi ) 
\right. \right.
+ 
\left( \frac{1}{\varepsilon_{UV}} - \frac{1}{\varepsilon_{IR}} \right)
\left( \frac{\xi}{1 - \xi} \right)_+
\nonumber\\   
&&
\!\!\!- \left.   
\left( \frac{1}{2\varepsilon_{IR} } 
+ L \right)\! \right\}\!
\langle 
\bar{q}(\xi tn) \Slash{n}
\gamma_5 h_v (0) 
\rangle  
-
\left.
   t\left( \frac{1}{\varepsilon_{IR}} + 2 L- 1 - \xi  \right)\!
\langle 
\bar{q}(\xi tn)v \cdot \overleftarrow{D}  \Slash{n}
\gamma_5 h_v (0)
\rangle  
\right] \!\!+\! \cdots,
\label{eq2}
\end{eqnarray}
in $D=4-2\varepsilon$ dimensions and Feynman gauge,
where 
$L\equiv \ln\left[i(t-i0) \mu e^{\gamma_E}\right]$
with the ${\overline{\rm MS}}$ scale $\mu$ and the Euler constant $\gamma_E$.
The ``vertex-type'' correction 
that connects the light-like Wilson line and $\bar{q}(tn)$
in (\ref{eq1}) 
is associated with only the massless degrees of freedom and
yields
the 
scaleless loop-integral that gives
the term 
with the ``canceling'' UV and IR poles, $1/\varepsilon_{UV}-1/\varepsilon_{IR}$,
and with
the ``plus''-distribution $(\xi/(1-\xi) )_+$ as the splitting function;
this term 
is identical to
the corresponding correction 
for the case of 
the pion LCDA. 
The other terms in (\ref{eq2})
have 
``non-canceling'' UV and IR poles:
another vertex-type correction around a ``cusp'' between the two Wilson lines,
the light-like Wilson line of (\ref{eq1}) and the time-like Wilson line from 
$h_v(0)={\rm P}\exp [ig\int_{-\infty}^0d\lambda v\cdot A(\lambda v)]h_v(-\infty v)$,
gives the 
terms proportional to 
$\delta (1-\xi)$, which contain
the double as well as single UV pole, corresponding to the cusp singularity~\cite{Lange:2003ff}.
The ``ladder-type'' correction, connecting the two quark fields in (\ref{eq1}), gives
all the remaining terms in (\ref{eq2}), which contain
the IR poles and are associated with not only the bilocal operator in (\ref{eq1}), 
but also the higher dimensional operators;
the ellipses in (\ref{eq2}) are expressed by the operators
involving 
two or more additional
covariant derivatives.

The renormalized LCDA is obtained by subtracting the UV poles from (\ref{eq2}) 
with the trivial quark self-energy corrections complemented.
Here, the term with the plus-distribution $(\xi/(1-\xi) )_+$ 
is analytic (Taylor expandable) at $t=0$, similar to the pion LCDA, but
the other terms
are not
analytic
due to the presence of logarithms $L$, $L^2$~\cite{Lange:2003ff,Braun:2003wx}.
In particular, the nontrivial dependence of the latter terms on $t\mu$ through $L$
implies that the scale $\sim 1/t$ separates the UV and IR regions.
Thus, we have to use the operator product expansion (OPE) to
treat the different UV and IR behaviors simultaneously:
the coefficient functions absorb all the singular logarithms,
while, for the local operators to absorb the IR poles, we have to take into account many 
higher dimensional operators.
Such OPE with local operators is useful when the separation $t$
is less than the typical distance scale of quantum fluctuation,
i.e., when $t\lesssim 1/\mu$.
We note that an OPE for the $B$-meson LCDA (\ref{eq1}) 
was discussed 
in \cite{Lee:2005gza},
taking into account the local operators of dimension 
$d \le$ 4
and 
the NLO ($O(\alpha_s)$)
corrections to the corresponding Wilson coefficients 
in a 
``cutoff scheme'', where
an additional momentum cutoff $\Lambda_{UV}$ ($\gg \Lambda_{\rm QCD}$) 
was introduced, 
and the OPE, in powers of $1/\Lambda_{UV}$, was derived for the regularized moments,
$M_j = \int_0^{\Lambda_{UV}} d\omega  \omega^j 
\phi_+ (\omega, \mu)$,
in particular, for the first two moments with $j=0, 1$; note,
$M_j \rightarrow \infty$ as $\Lambda_{UV} \rightarrow \infty$~\cite{Grozin:1997pq}.
Here, we derive the OPE for 
(\ref{eq1}),
taking into account the local operators of dimension 
$d \le$ 5
and calculating 
the corresponding Wilson coefficients
at NLO accuracy.
Following the discussion above, we carry out the calculation for $t\lesssim 1/\mu$
in the coordinate space and
in the $\overline{\rm MS}$ scheme, 
so that there is no need to introduce any additional cutoff.

The most complicated task
is the reorganization of contributions from 
(many) Feynman diagrams 
in terms of the
matrix element
of 
gauge-invariant operators including higher dimensional operators, 
in particular, 
the three-body operators of dimension 5,
such as $\bar{q}G_{\alpha\beta}\Slash{n}  \gamma_5 h_v$ with
the field strength tensor $G_{\alpha\beta}$~\cite{Grozin:1997pq,KKQT}.
To derive the NLO Wilson coefficients associated with such operators,
we have to compute the 1-loop diagrams for the 3-point function,
as well as those for the 2-point function as in (\ref{eq2}),
where the former diagrams are obtained by attaching the external gluon line
to the latter diagrams in all possible ways. 
We employ
the background 
field method~\cite{Abbott:1980hw},
where the background fields represent the nonperturbative long-distance 
degrees of freedom and satisfy the exact classical EOM.
We use the Fock-Schwinger gauge, $x^\mu  A_\mu^{(c)} (x)=0$,
for the background gluon field $A_\mu^{(c)}$.
This gauge condition 
is solved to give 
$A_\mu^{(c)}(x) = \int_0^1 {du} u x^\beta  G_{\beta \mu}^{(c)} (ux)$~\cite{Abbott:1980hw},
which allows us to reexpress each Feynman diagram in terms of the
matrix element of the operators associated with the field strength tensor.
Also, this ensures that the Wilson line in (\ref{eq1}), 
as well as the heavy-quark propagator,
does not couple directly to the background gluons
while a massless quark or gluon propagator
couples to them.
With the 
matching
in the $\overline{\rm MS}$ scheme,
we obtain~\cite{KT09}
the OPE,
\begin{equation}
\bar{q}(tn)
{\rm P}e^{ig\int_0^td\lambda n\cdot A(\lambda n)}
\Slash{n}\gamma_5h_v(0)
=C_1^{(3)}(t,\mu) {\cal O}_1^{(3)}(\mu) +\sum_{k=1}^2 C_k^{(4)}(t,\mu) {\cal O}_k^{(4)}(\mu)
+\sum_{k=1}^7 C_k^{(5)}(t,\mu) {\cal O}_k^{(5)}(\mu)\ ,
\label{eq3}
\end{equation}
where the summation is over a basis of local operators of dimension-$d$, ${\cal O}^{(d)}_k$
($k=1,2,\ldots$), 
defined as
${\cal O}^{(3)}_1\equiv \bar{q}n\hspace{-0.45em}/\gamma_5h_v$,
$\{ {\cal O}^{(4)}_k \}\equiv\{ \bar{q}(i n\cdot \overleftarrow{D})
n\hspace{-0.45em}/\gamma_5h_v$,
$\bar{q}(iv\cdot \overleftarrow{D})
n\hspace{-0.45em}/\gamma_5h_v\}$, 
and
$\{ {\cal O}^{(5)}_k \}\equiv\{ 
\bar{q}(in\cdot \overleftarrow{D})^2
n\hspace{-0.45em}/ \gamma_5 h_v$,
$\bar{q}(iv\cdot \overleftarrow{D})
(in\cdot \overleftarrow{D})n\hspace{-0.45em}
/\gamma_5 h_v$,
$\bar{q}(iv\cdot \overleftarrow{D})^2
n\hspace{-0.45em}/\gamma_5 h_v$,
$\bar{q}igG_{\alpha\beta}v^\alpha n^\beta n\hspace{-0.45em}/
\gamma_5h_v$,
$\bar{q}igG_{\alpha\beta}\gamma^{\alpha}n^{\beta}
\bar{n}\hspace{-0.45em}/\gamma_5h_v$,
$\bar{q}igG_{\alpha\beta}\gamma^\alpha v^\beta
\bar{n}\hspace{-0.45em}/
\gamma_5h_v$,
$\bar{q}gG_{\alpha\beta}\sigma^{\alpha\beta}
n\hspace{-0.45em}/\gamma_5h_v\}$,
with another light-like vector, $\bar{n}^2=0$,
as $v_\mu=(n_\mu+\bar{n}_\mu)/2$. 
The NLO Wilson coefficients 
are obtained as 
\begin{eqnarray}
C_1^{(3)}(t,\mu)&& \!\!=
1-  \frac{\alpha_sC_F}{4\pi}
\left(2L^2+2L+ \frac{5\pi^2}{12}\right)\ , \;\;\;
C_1^{(4)}(t,\mu)= -it
\left[
1- \frac{\alpha_sC_F}{4\pi}
\left(2L^2+L+\frac{5\pi^2}{12}\right)
\right]\ ,
\nonumber\\
C_2^{(4)}(t,\mu)&& \!\!= 
\frac{it\alpha_sC_F}{4\pi}\left(4L-3\right)\ , \;\;\;
C_1^{(5)}(t,\mu)= -\frac{t^2}{2}
\left[1- \frac{\alpha_sC_F}{4\pi}
\left(
2L^2+\frac{2}{3}L+\frac{5 \pi^2}{12}
\right)\right]\ , 
\label{wc}
\end{eqnarray}
and, for the explicit form of $C_{2}^{(5)}(t,\mu), C_{3}^{(5)}(t,\mu), \ldots, C_{7}^{(5)}(t,\mu)$,  
we refer the readers to \cite{KT09}.
Here and below, 
$\mu$ is the $\overline{\rm MS}$ scale, and $\alpha_s \equiv \alpha_s(\mu)$.
The double logarithm $L^2$ in the coefficient functions
originates from the cusp singularity (see (\ref{eq2})).
The 1-loop corrections 
for the 2-point function
induce all of the above ten operators 
using the EOM,
while those
for the 3-point function 
induce only 
${\cal O}^{(5)}_{4,5,6,7}$
associated with the field-strength tensor; as a result, 
the 
coefficients $C_{4,5,6,7}^{(5)}(t,\mu)$ 
involve the terms proportional to 
the color factor $C_G=N_c$ as well as to $C_F$~\cite{KT09}. 

Taking the matrix element $\langle \cdots \rangle \equiv \langle 0| \cdots|\bar{B}(v) \rangle$
of (\ref{eq3}), we can derive the OPE form of 
the $B$-meson LCDA (\ref{eq1}).
The matrix elements of the local operators in (\ref{eq3})
are known to be related to a few nonperturbative parameters
in the HQET, using the EOM and heavy-quark symmetry as demonstrated in
\cite{Grozin:1997pq,KKQT}:
$\langle {\cal O}^{(4)}_1 
\rangle= 4iF(\mu)  \bar{\Lambda}/3$, 
$\langle {\cal O}^{(4)}_2 
\rangle
= iF(\mu) \bar{\Lambda}$, 
with $F$ of (\ref{eq1})
and $\bar{\Lambda}=m_B -m_b$, representing the mass difference
between the $B$-meson and $b$-quark,
and all seven matrix elements $\langle {\cal O}^{(5)}_k  \rangle$ for the dimension-5 
operators 
can be expressed by $F$,
$\bar{\Lambda}$ and two additional HQET parameters $\lambda_E$ and $\lambda_H$,
which are associated with the
chromoelectric and chromomagnetic fields inside the $B$ meson as
$\langle \bar{q}g\bm{E}\cdot \bm{\alpha}
\gamma_5 h_v \rangle
=F(\mu)\lambda_E^2(\mu)$ and
$\langle \bar{q}g\bm{H}\cdot \bm{\sigma}
\gamma_5h_v \rangle
=i F(\mu)
\lambda_H^2(\mu)$, respectively, in the rest frame where $v=(1,{\bf 0})$.
As a result, we obtain~\cite{KT09} the OPE form for the LCDA (\ref{eq1}),
\begin{eqnarray}
&&\!\!\!\!
\tilde{\phi}_+(t,\mu)
=
1- \frac{\alpha_s C_F}{4\pi}
\left(2L^2+2L+\frac{5 \pi^2}{12}\right)
-it\frac{4\bar{\Lambda}}{3} 
\left[1- \frac{\alpha_s C_F}{4\pi}
\left(2L^2+4L-\frac{9}{4}+\frac{5\pi^2}{12} \right)
\right]
\nonumber\\
&&\!\!\!\!
-t^2 \bar{\Lambda}^2 \!\! \left[
1\! - \! \frac{\alpha_sC_F}{4\pi}\!\!
\left(2L^2+\frac{16}{3}L-\frac{35}{9}
+\frac{5\pi^2}{12} \right)\!
\right]
\!\!- \! \frac{t^2\lambda_E^2(\mu)}{3}\!
\left[1\! - \! \frac{\alpha_sC_F}{4\pi}\!
\left( \! 2L^2+2L-\frac{2}{3}
+\frac{5\pi^2}{12} \! \right)
\right.
\nonumber\\
&&\;\;
\left.
+ 
\frac{\alpha_sC_G}{4\pi}
\left(\frac{3}{4}L-\frac{1}{2}\right)
\right]
-\frac{t^2\lambda_H^2(\mu)}{6} 
\left[1- \frac{\alpha_sC_F}{4\pi}
\left(2L^2+\frac{2}{3}
+\frac{5\pi^2}{12} \right)
-\frac{\alpha_s C_G}{8\pi}
\left(L-1\right)
\right]\ ,
\label{eq4}
\end{eqnarray}
which takes into account the Wilson coefficients 
to $O(\alpha_s)$ and a complete set of the local operators of dimension 
$d \le$ 5.
Fourier transforming to the momentum 
representation
and taking the first two ($j=0,1$) regularized-moments,
$M_j=\int_0^{\Lambda_{UV}} d\omega  \omega^j \phi_+ (\omega, \mu)$,
the contributions from the first line in (\ref{eq4}), associated with matrix elements of
the dimension-3 and -4 operators, coincide completely with
the result obtained in \cite{Lee:2005gza}.
The second and third lines in (\ref{eq4}) are
generated from the dimension-5 operators.
Our OPE result (\ref{eq4}) ``merges'' the UV~\cite{Lange:2003ff} and IR structures~\cite{KKQT} 
peculiar to the $B$-meson LCDA, 
so that it embodies novel behaviors that are completely different from those
of the pion LCDA:
$\mu$ and $t$ are strongly correlated due to the
logarithmic contributions, $L= \ln\left[i(t-i0) \mu e^{\gamma_E}\right]$,
from radiative corrections,
so that the DA is not Taylor expandable about $t=0$, which in turn implies
the UV divergence in the moments~\cite{Grozin:1997pq,Braun:2003wx,Lee:2005gza},
$M_j \rightarrow \infty$ as $\Lambda_{UV} \rightarrow \infty$.
The DA receives the contributions from (many) higher dimensional operators,
in particular, from those associated with the long-distance gluon fields inside
the $B$-meson. 
It is instructive to draw a comparison 
with the previous results,
concerning
UV or IR structure:
one can prove~\cite{KT09} that (\ref{eq4})
satisfies the renormalization group 
equation for (\ref{eq1}), 
which is governed by
the evolution kernel~\cite{Lange:2003ff} determined
by the (single) UV poles in (\ref{eq2}).
On the other hand, (\ref{eq4}) reveals that the solution
of the EOM constraints for (\ref{eq1}), which was obtained in \cite{KKQT}, 
is subject to additional effects from radiative corrections,
see \cite{KT09} for the detail (see also 
\cite{Braun:2003wx}).
Such corrections to
the EOM constraints 
at order $\alpha_s$
in perturbation theory
is peculiar to the heavy-meson LCDAs in the HQET and does not arise 
for the case of the (higher twist) LCDAs 
for the light mesons, $\pi, \rho$, etc.~\cite{Braun:1990iv}.

Our OPE form (\ref{eq4}) allows us to parameterize all nonperturbative contributions 
in the $B$-meson LCDA (\ref{eq1}) 
by three HQET parameters,
$\bar{\Lambda}$,
$\lambda_E$ and $\lambda_H$,
and gives a model-independent description 
of the $B$-meson LCDA 
when $t \lesssim 1/\mu$ ($\leq 1/\Lambda_{\rm QCD}$),
taking into account the UV and IR structures simultaneously.
Here, we evaluate (\ref{eq4}) at the scale $\mu =1$~GeV:
$\bar{\Lambda}=m_B - m_b$ in (\ref{eq4}) is defined by
the $b$-quark pole mass $m_b$. 
Following \cite{Lee:2005gza}, 
we eliminate $\bar{\Lambda}$ in favor of a short-distance parameter, $\bar{\Lambda}_{DA}$,
free from IR renormalon ambiguities and written as
$\bar{\Lambda}
   = \bar{\Lambda}_{DA}(\mu)
    \left[ 1 + (7/16\pi) C_F\alpha_s
\right] 
-( 9/8\pi)\mu C_F\alpha_s$,
to one-loop accuracy;
$\bar{\Lambda}_{DA}(\mu)$ can be related to another short-distance mass parameter
whose value is extracted from analysis of the spectra in inclusive decays $B\to X_s\gamma$ 
and $B\to X_u l\,\nu$, leading to
$\bar{\Lambda}_{DA}(\mu =1~{\rm GeV})\simeq 0.52$~GeV~\cite{Lee:2005gza}. 
For the novel 
parameters associated with the
dimension-5 operators, we use the central values of
$\lambda_E^2 (\mu) = 0.11 \pm 0.06~{\rm GeV}^2$,
$\lambda_H^2 (\mu) = 0.18 \pm 0.07~{\rm GeV}^2$,
at $\mu =1$~GeV,
which were 
obtained by QCD sum rules~\cite{Grozin:1997pq}; no other estimate
exists for $\lambda_{E}$ or $\lambda_{H}$.
We calculate
(\ref{eq4}) 
for imaginary LC separation, performing the Wick rotation
$t\rightarrow -i\tau$~\cite{Grozin:1997pq,Braun:2003wx}.

\begin{figure}
\includegraphics[width=0.51\textwidth]{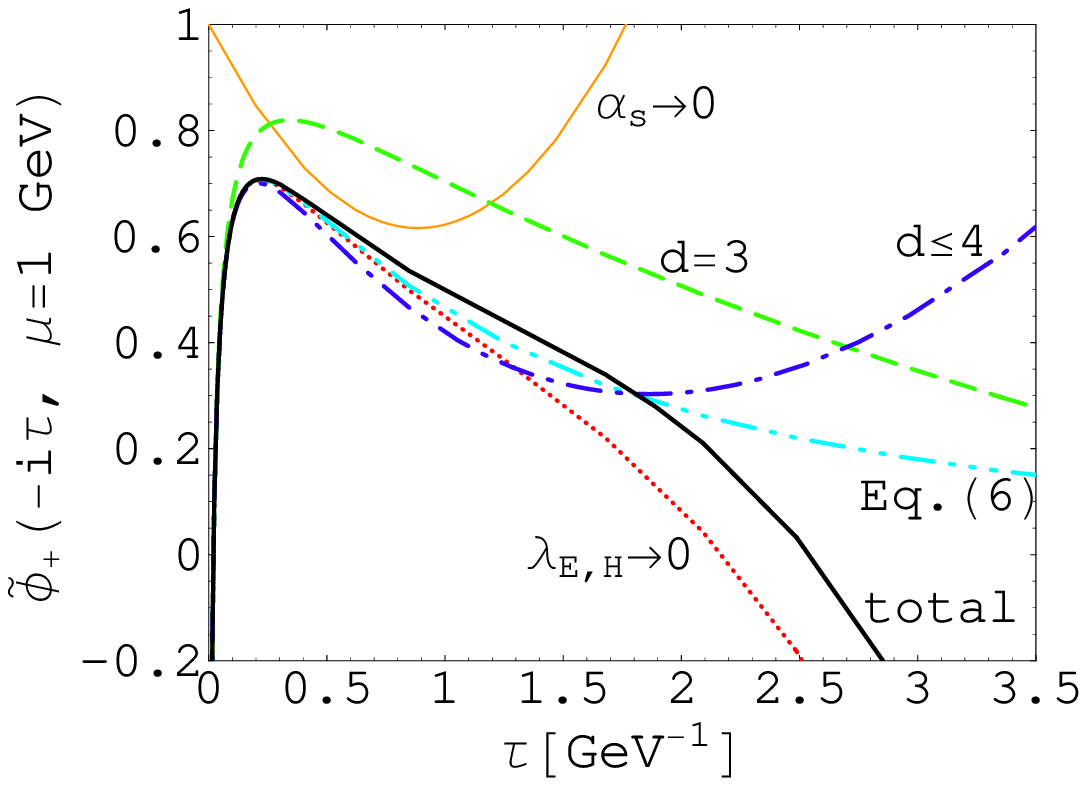}
\includegraphics[width=0.50\textwidth]{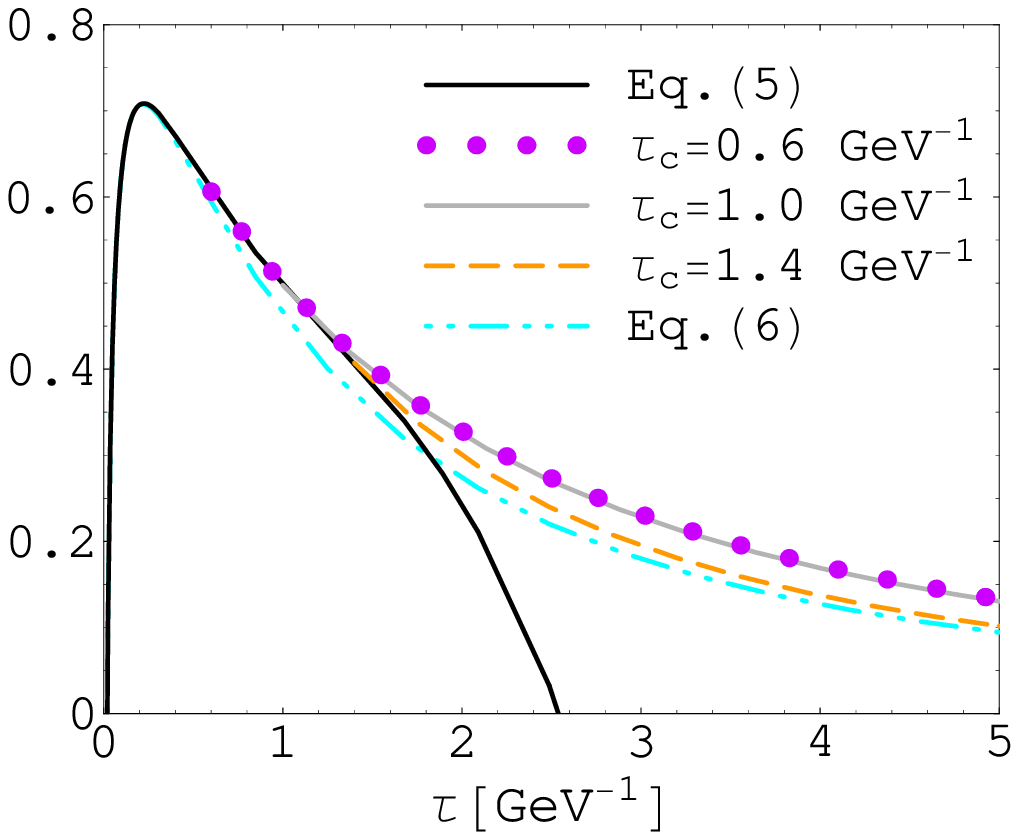}
\caption{The $B$-meson LCDA 
at $\mu=1$~GeV
using the OPE (left) and its continuation with a model 
(right).}
\end{figure}
The results for $\tilde{\phi}_{+}(-i\tau, \mu=1~{\rm GeV})$ 
using (\ref{eq4}) are shown as
a function of $\tau$ in the LHS of Fig.~1~\cite{KT09}:
the wide-solid curve 
shows the whole contributions 
of (\ref{eq4}),
while the narrow-solid curve shows the result
for $\alpha_s \rightarrow 0$; 
the NLO perturbative corrections are at the 10-30\% level for moderate $\tau$ 
of order 1~GeV$^{-1} \sim 1/\mu$,
while they are very large for $\tau \rightarrow 0$ because of singular logarithms $L^2$ and $L$.
The dashed and dot-dashed curves show the contributions
of the first two terms and the first line in (\ref{eq4}), respectively,
associated with the operators of dimension 
$d =3$ and $d \le 4$,
while the dotted curve gives the results of (\ref{eq4}) 
when $\lambda_E=\lambda_H = 0$.
For moderate $\tau$, the contributions from the dimension-4 operators suppress the DA by
30-40\%, but
the dimension-5 operators, in contrast, lead to enhancement 
by 10-20\%
with significant effects from $\lambda_E$ and $\lambda_H$.          
Our $B$-meson LCDA (\ref{eq4})
indeed works up to moderate LC distances $\tau$,
where the hierarchy among the dashed, dot-dashed, and wide-solid curves
demonstrates convergence of the OPE (\ref{eq3}).

The two-dot-dashed curve in the LHS of Fig.~1 shows
the behavior of the two-component ansatz
by Lee and Neubert~\cite{Lee:2005gza}, 
which is given in momentum space as
\begin{equation}
   \phi_+^{\rm LN}(\omega,\mu)
   = N\,\frac{\omega}{\omega_0^2}\,
    e^{-\omega/\omega_0} + \theta(\omega-\omega_t)\,
    \frac{C_F\alpha_s}{\pi\omega} 
 \left[ \left( \frac12 - \ln\frac{\omega}{\mu} \right)
    + \frac{4\bar\Lambda_{DA}}{3\omega}
    \left( 2 - \ln\frac{\omega}{\mu} \right) \right]\ ,
\label{LNmodel}
\end{equation}
where the second term reproduces the correct asymptotic behavior of the DA (\ref{eq1})
for $\omega \gg \Lambda_{\rm QCD}$
and the first term represents the nonperturbative component 
modeled by an exponential form~\cite{Grozin:1997pq},
with $\omega_t = 2.33$~GeV, $N=0.963$, and $\omega_0 = 0.438$~GeV at 
$\mu=1$~GeV;
these parameters
are fixed by matching  
the first two ($j=0,1$) cut-moments 
$\int_0^{\Lambda_{UV}} d\omega  \omega^j 
\phi_+^{\rm LN} (\omega, \mu)$
with the OPE for the corresponding cut-moments $M_{0,1}$ derived in \cite{Lee:2005gza},
where the operators of dimension 
$d \le$ 4
and the corresponding Wilson coefficients at NLO are taken into account.
For $\tau \lesssim 1$~GeV$^{-1}$,
the Lee-Neubert ansatz (\ref{LNmodel}) shows behavior similar 
to (\ref{eq4}) with $\lambda_E=\lambda_H = 0$ substituted;
note that the first term of (\ref{LNmodel}) 
produces particular
contributions associated with the operators of dimension-5 and higher.

For $\tau \gg 1$~GeV$^{-1}$,
the contributions associated with higher-dimensional operators
become important,
and the OPE diverges (see (\ref{eq4}) and Fig.~1); thus, one has to rely on a certain model for 
the large $\tau$ behavior and connect the model-independent descriptions at small and moderate $\tau$
to that model.
The results in Fig.~1 suggest the possibility of connecting the behavior 
for $\tau \le \tau_c$ ($\tau_c \sim 1$~GeV$^{-1}$)
given by our OPE form  (\ref{eq4}) 
to that for $\tau \ge \tau_c$, given by the coordinate-space representation
of the first term of (\ref{LNmodel}),
$\int_0^\infty d\omega e^{- \omega\tau}\left( N\omega/\omega_0^2\right)
    e^{-\omega/\omega_0}=
N/\left( \tau \omega_0 +1\right)^2$.
Here, $N$ and $\omega_0$ 
can be determined such that
both the resulting total DA $\tilde{\phi}_+(-i\tau,\mu)$ and its derivative
$\partial \tilde{\phi}_+(-i\tau,\mu)/\partial \tau$ are continuous
at 
$\tau=\tau_c$.
In the LHS of
Table~1, 
we show~\cite{KT09} the values of $N$ and $\omega_0$ obtained by solving 
the corresponding conditions of the continuity
for $\mu=1$~GeV. (The RHS of Table~1
shows 
the results that 
would be obtained by solving the similar continuity conditions 
with $\lambda_E=\lambda_H=0$.)
\begin{table}
\begin{tabular}{|c|c|c|c||c|c|c|}
\hline
&\multicolumn{3}{|c||}{$\lambda_E^2=0.11$~{\small GeV$^2$},~~$\lambda_H^2=0.18$~{\small GeV$^2$}}
&
\multicolumn{3}{|c|}{$\lambda_E^2=\lambda_H^2=0$}\\
\hline
$\tau_c$~{\small [GeV$^{-1}$]}& $N$ & $\omega_0$~{\small [GeV]} 
& $\lambda_B^{-1}$~{\small [GeV$^{-1}$]} 
& $N$ & $\omega_0$~{\small [GeV]} 
& $\lambda_B^{-1}$~{\small [GeV$^{-1}$]} \\
\hline
0.4  & 0.816 & 0.257 & $ 3.11\, \  ( 0.23  + 2.88 )$ 
     & 0.832 & 0.301 & $ 2.69\, \  ( 0.23  + 2.46 )$ \\
0.6  & 0.850 & 0.306 & $ 2.70\, \  ( 0.35  + 2.35 )$
     & 0.899 & 0.394 & $ 2.19\, \  ( 0.35 + 1.84 )$ \\
0.8  & 0.852 & 0.308 & $ 2.69\, \  ( 0.47  + 2.22 )$ 
     & 0.966 & 0.461 & $ 1.99\, \  ( 0.46 + 1.53 )$ \\
1.0  & 0.858 & 0.313 & $ 2.66\, \  ( 0.58  + 2.08  )$
     & 1.11 & 0.572 & $ 1.79\, \  ( 0.56 + 1.23 )$ \\
1.2  & 0.910 & 0.349 & $ 2.51\, \  ( 0.67  + 1.84 )$
     & 1.55 & 0.839 & $ 1.56\, \  ( 0.64 + 0.92 )$ \\
1.4  & 1.09  & 0.456 & $ 2.22\, \  ( 0.76  + 1.46 )$
     & 4.43 & 1.95 & $ 1.32\, \  ( 0.71 + 0.61 )$ \\
1.6  & 1.81  & 0.777 & $ 1.87\, \  ( 0.83  + 1.04 )$
     & 9.82 & $-4.55$ & $1.11\, \  ( 0.77 + 0.34 )$ \\
\hline
\end{tabular}
\caption{Parameters of the model function $N/\left( \tau \omega_0 +1\right)^2$
for $\tau \ge \tau_c$
with different values of $\tau_c$,
and
the results of 
the inverse moment 
$\lambda_B^{-1}(\mu)$
at $\mu=1$~GeV,
with the first and second 
numbers in the parentheses denoting the contributions from the first and the second terms
in the RHS of (7).}
\end{table}
In the RHS of Fig.~1,
the wide-solid and two-dot-dashed curves
are same as those
in the LHS,
and 
the dotted, 
solid-gray,
and dashed curves show the behavior of the above model function 
$N/\left( \tau \omega_0 +1\right)^2$
for $\tau \ge \tau_c$ with 
$\tau_c=0.6$, $1.0$, and $1.4$~GeV$^{-1}$, 
respectively,
using the corresponding values of $N$ and $\omega_0$ in the LHS of
Table~1;
these three curves behave as 
$\sim N/(\omega_0^2 \tau^2)$ at large $\tau$,
with 
larger $N/\omega_0^2$ 
than those of
(\ref{LNmodel}) and the RHS of Table~1.
Indeed, 
we can show that
$N/\omega_0^2=(9/4 \bar{\Lambda}_{DA}^2)\left\{1+\tau_c \bar{\Lambda}_{DA}
\left[ \lambda_E^2/ \bar{\Lambda}_{DA}^2+\lambda_H^2/ (2\bar{\Lambda}_{DA}^2)
-1\right]\right\} +\cdots$, using the continuity 
of $\tilde{\phi}_+(-i\tau,\mu)$, $\partial \tilde{\phi}_+(-i\tau,\mu)/\partial \tau$ 
at $\tau=\tau_c$, 
and thus the contributions of 
$\lambda_E$ and $\lambda_H$ enhance $N/\omega_0^2$.

Using these results, 
we calculate the first inverse moment of the LCDA,
\begin{equation}
\lambda_B^{-1}(\mu) = \int_0^\infty d\omega
    \frac{\phi_+(\omega,\mu)}{\omega} = \int_0^{\tau_c} d\tau 
\tilde{\phi}_{+}(-i\tau, \mu)+ \int_{\tau_c}^\infty d\tau 
\tilde{\phi}_{+}(-i\tau, \mu)\ ,
\label{lambdaB}
\end{equation}
which is of particular interest for the QCD description of exclusive $B$-meson decays.
We substitute (\ref{eq4}) and the model function,
$N/\left( \tau \omega_0 +1\right)^2$,
into the first and the second terms 
in the RHS, respectively, and the results 
are shown in Table~1
for each value of $\tau_c$~\cite{KT09}.
The ``stable'' behavior observed for
$0.6~{\rm GeV}^{-1} \lesssim \tau_c \lesssim 1~{\rm GeV}^{-1}$ in the LHS of Table~1 
and in the RHS of Fig.~1 suggests that $\lambda_B^{-1}(\mu=1~{\rm GeV})\simeq 2.7$~GeV$^{-1}$,
i.e., $\lambda_B(\mu=1~{\rm GeV})\simeq 0.37$~GeV.
This value of $\lambda_B$ is somewhat smaller 
than the previous estimates that include nonperturbative and/or perturbative
QCD corrections~\cite{Braun:2003wx,Lee:2005gza,Khodjamirian:2005ea}
(e.g., (\ref{LNmodel}) 
gives $\lambda_B(\mu=1~{\rm GeV}) \simeq 0.48$~GeV).
A value of $\lambda_B$ that is as small as our value 
was adopted in \cite{Beneke:2000ry}.
Note that in the RHS of Table~1 with $\lambda_{E,H} = 0$, 
the stable behavior is not seen as clearly as in the LHS,
and $\lambda_B$ assumes larger values than in the latter.
These results demonstrate that 
the novel
HQET parameters, $\lambda_E$ and $\lambda_H$, 
associated with the dimension-5 quark-antiquark-gluon operators, 
could lead to smaller value  
of $\lambda_B$. 
In particular, 
using 
the values 
$\lambda_E^2 = 0.17~{\rm GeV}^2$,
$\lambda_H^2 = 0.25~{\rm GeV}^2$,
which correspond to 
their upper bound from the QCD sum rule estimate at $\mu=1$~GeV~\cite{Grozin:1997pq},
we find that 
the wide-solid curve in Fig.~1 
becomes further
enhanced 
in the moderate $\tau$ region, so that (\ref{lambdaB}) gives
$\lambda_B(\mu=1~{\rm GeV})
\sim 0.2$~GeV or smaller.

To summarize, 
we have derived the OPE 
that embodies both the notorious UV and IR behaviors of the $B$-meson LCDA, 
including all contributions from the local operators of dimension $d\leq 5$ and 
the corresponding Wilson coefficients at NLO accuracy.
This OPE 
provides us with 
the most accurate
description of the $B$-meson LCDA for distances less than $\sim 1/\Lambda_{\rm QCD}$. 
We have also used the model-independent behaviors from our OPE
to constrain the long-distance behavior of the LCDA and estimate 
the first inverse moment $\lambda_B^{-1}$.
The results demonstrated the 
impact of the novel HQET parameters, associated with 
the dimension-5 quark-antiquark-gluon 
operators.

\end{document}